# A crystal-field route to THz-driven magnetization

T. Zalewski,[1,2,*] M.S. Mrudul[3], Y. Lee[4], M. Weissenhofer[3], A.V. Boris[4], P.M. Oppeneer[3], A. Kirilyuk[1,2] and C.S. Davies[1,2,*]

[1] *HFML-FELIX, Toernooiveld 7, 6525 ED Nijmegen, The Netherlands*

[2] *Radboud University, Institute for Molecules and Materials, Heyendaalseweg 135, 6525 AJ Nijmegen, The Netherlands*

[3] *Department of Physics and Astronomy, Uppsala University, P.O. Box 516, SE-75120 Uppsala, Sweden*

[4] *Max Planck Institute for Solid State Research, Heisenbergstrasse 1, 70569 Stuttgart, Germany*

* Corresponding authors: tomasz.zalewski@ru.nl and carl.davies@ru.nl

## ABSTRACT

Light carries angular momentum, but the microscopic pathways that transform it into magnetization remain elusive. Here we establish that crystal-field excitations, historically viewed primarily as equilibrium spectroscopic fingerprints of localized 4*f* electrons, constitute an active microscopic route through which circularly-polarized terahertz (THz) light creates magnetic polarization. Using wavelength-selective ultrafast Faraday spectroscopy on the paramagnetic insulator $CeF_3$, we show that resonant excitation of localized 4*f* crystal-field transitions generates a helicity-dependent magnetization that survives for up to about 100 ps. Most strikingly, while the optical helicity is held fixed, the THz-driven response reverses sign as the excitation wavelength is tuned across the crystal-field resonance. The resulting dispersive spectral response follows the crystal-field excitation spectrum rather than that of optical phonons, and is captured by resonant electronic theory of the inverse Faraday effect. Our results identify crystal-field excitations as a previously unrecognized dynamical reservoir for optical angular momentum and broaden the microscopic pathways through which THz light can create and manipulate magnetic states.

## MAIN TEXT

Transferring angular momentum from light to matter provides a powerful route for generating and controlling magnetization [1]-[3]. The search for microscopic degrees of freedom capable of mediating this flow has intensified dramatically in recent years, with circularly-polarized phonons emerging as a prominent example through their roles in the ultrafast Einstein–de Haas and Barnett effects [4]-[9]. In principle, the ideal reservoir addressable by optical angular momentum would combine efficient optical accessibility with an immediate microscopic link between angular momentum and magnetic moment, yielding a direct pathway from optical helicity to magnetization. Surprisingly though, excitations that unite these attributes within a single microscopic degree of freedom are remarkably rare.

Against this backdrop, rare-earth crystal-field excitations offer a compelling platform [10]. The localized 4*f* electrons of rare-earth ions occupy discrete crystal-field states that inherit well-defined angular momentum and magnetic moments from the underlying atomic multiplets. Because both angular momentum and magnetic moment reside in the same microscopic states, crystal-field transitions provide an unusually direct pathway from optical angular momentum to magnetization, while their spectral isolation facilitates selective resonant excitation. Yet, despite this unique combination of properties, crystal-field excitations have traditionally been regarded as static ingredients of magnetic anisotropy and equilibrium thermodynamic behavior rather than active participants in nonequilibrium dynamics [11]. Whether they can serve as microscopic reservoirs for optical angular momentum has therefore remained largely unexplored.

Here, we show that this possibility is realized in $CeF_3$, a paramagnetic dielectric whose low-lying $Ce^{3+}$ 4*f* crystal-field transitions lie in the far-infrared range. By tuning circularly polarized THz pulses through these transitions and monitoring the response with time-resolved Faraday rotation, we observe the ultrafast creation of helicity-odd spin polarization that survives for more than ≈100 ps. Most importantly, even when holding the optical helicity fixed, tuning through a crystal-field resonance also reverses the sign of the induced magnetization. This inversion shows that the magnetic response is governed by the resonant 4*f* crystal-field transition, rather than simply by absorbed energy or phonon excitation. A quantum theory of the inverse Faraday effect based on optically-selected crystal-field transitions explains this behavior, revealing a crystal-field-mediated route by which circularly polarized THz light can generate non-equilibrium magnetization in a paramagnet.

## Helicity- and wavelength-selective magnetization

To directly examine whether localized crystal-field excitations can translate optical angular momentum into magnetization, we concentrate our study on the paramagnetic insulator $CeF_3$. Its lack of magnetic order in equilibrium makes $CeF_3$ a particularly clean system for isolating optically-generated magnetization. $CeF_3$ crystallizes in the trigonal tysonite structure, where localized $Ce^{3+}$ $4f^1$ electrons experience an electrostatic crystal field generated by the surrounding fluorine ions [12]. The $Ce^{3+}$ $^2F_{5/2}$ ground-state multiplet is consequently split into three Kramers doublets [13], with the first and second excited doublets lying at approximately 140 and 290 $cm^{-1}$ (71 and 34 μm), respectively [14]. These far-infrared crystal-field states are mixed superpositions of atomic $|J, m_J\rangle$ states $(J = \frac{5}{2}, m_J = \pm\frac{5}{2}, \pm\frac{3}{2}, \pm\frac{1}{2})$ carrying finite angular momentum and magnetic moments, thereby offering a direct microscopic link between optical excitation and magnetization. In equilibrium, time-reversal symmetry enforces equal populations of the two members of each Kramers doublet, yielding zero net magnetization despite the finite moments carried by the individual states.

To drive crystal-field transitions at resonance, we excite $CeF_3$ at 4 K with intense transform-limited THz transients generated by a free-electron laser [15]. The pulses have narrow spectral bandwidth (see Supplemental Data), facilitating selective excitation of individual transitions. Moreover, a wavelength-independent silicon phase retarder ensures circular polarization with preserved helicity across the entire far-infrared spectral range [16]. The circularly-polarized pump pulses are focused at normal incidence onto the $CeF_3$ crystal, defining the optical axis along which the induced magnetization is generated. The resulting out-of-plane magnetization dynamics are monitored using time-resolved Faraday rotation of a synchronized 150-fs probe pulse at 515 nm.

Figure 1A shows representative time traces obtained for excitation wavelengths of 74 and 88 μm, with both signals exhibiting an ultrafast rise followed by exponential decay across a timescale of ≈100 ps. Remarkably, while the helicity of the THz excitation is held fixed, the induced Faraday rotation reverses sign when the excitation wavelength is tuned from near-resonance to the long-wavelength side of the transition. Here, positive and negative magneto-optical Faraday rotations correspond to magnetization oriented parallel and antiparallel to the wavevector of the far-infrared pump pulse, respectively. The direction of the light-induced magnetization is therefore clearly governed by both the helicity of the optical pulse and its spectral detuning relative to the crystal-field transition. The role of optical angular momentum in the effect is confirmed by the exact inversion of the transient Faraday rotation upon reversing the optical helicity (Fig. 1B).

Beyond controlling the magnetization sign, the THz excitation also produces a significantly large magnetic response. Using the equilibrium magneto-optical calibration previously established for $CeF_3$ [6], our observed maximum transient Faraday rotation of ≈40 mdeg corresponds to the magnetization induced by an external magnetic field of order 10 T. This equivalent field scale does not represent a real applied magnetic field [17]-[18], but instead reflects the internal polarization of the $Ce^{3+}$ 4*f* moments generated through optical angular-momentum transfer.

**Spectroscopic signatures of crystal-field magnetization**

The pronounced wavelength- and helicity-dependence of the light-induced magnetization motivates a systematic investigation of its spectral response. To explore this, we fixed the pump-probe delay at the maximum of the transient response observed at 74 μm, and continuously scanned the excitation wavelength across a sixfold range spanning 20 to 120 μm. The resulting pump-induced polarization rotation, shown in Fig. 2A (black points), reveals a strongly structured response, with the largest signal occurring near 75 μm. Remarkably, tuning the excitation wavelength through this region reverses the direction of the induced magnetization, resulting in a dispersive spectral response that crosses zero at 82 μm before reaching a maximum with the opposite sign at 88 μm. A second clear helicity-dependent feature emerges near 34 μm, albeit without a comparable inversion of sign. Reversing the pump helicity inverts the entire spectral response (red points), demonstrating deterministic control of the magnetization direction by the optical angular momentum. In contrast, excitation with linearly polarized light produces no measurable response across the full spectral range (blue points), establishing that the angular momentum carried by the incident THz radiation is essential for generating the magnetic state.

To elucidate the origin of the THz-driven magnetization, we compare the measured magneto-optical spectrum with the far-infrared optical spectrum of $CeF_3$ obtained at 12 K using synchrotron-based infrared ellipsometry [19]. The corresponding imaginary part of the dielectric function $\varepsilon_2(\omega)$ is shown in Fig. 2B, with the real part $\varepsilon_1(\omega)$, optical loss function and absorption given in Supplementary Materials. The optical response is dominated by a series of eleven infrared-active doubly-degenerate $E_u$ optical phonons, consistent with group-theoretical vibrational analysis for the trigonal $P\bar{3}c1$ space group [20]. These modes nominally support circular ionic motion and have accordingly been proposed as mediators of optical angular-momentum transfer in $CeF_3$ [6], although the microscopic pathway connecting axial lattice vibrations to macroscopic magnetization remains unresolved [21]-[24]. In stark contrast to expectations based on a purely phononic mechanism, the helicity-dependent magnetization is maximized near the crystal-field transition positioned at ≈71 μm, in a spectral window where phonon absorption is very weak. Conversely, several of the strongest phonon resonances, immediately adjacent at 60 and 100 μm, correlate with little or no measurable response despite their much larger oscillator strengths. This points to crystal-field excitations as the primary microscopic reservoir underpinning the observed light-induced magnetization.

The microscopic origin of the response near 30 μm requires a more nuanced interpretation. While earlier work attributed light-induced magnetization in this spectral region to the excitation of circularly polarized phonons [6], the helicity-dependent magneto-optical response is also near the higher-energy crystal field transition at 34 µm. Moreover, the magnitude of this response does not simply follow the local population of the infrared-active phonon modes (see Supplementary Materials for a zoomed comparison). Instead, the strongest response occurs near the 34-µm crystal-field excitation, which is known to hybridize with nearby $E_{1g}$ Raman-active phonons to form vibronic modes [12] (see Supplementary Materials). Owing to the high density of both Raman- and infrared-active phonons in this spectral range, it is difficult to unambiguously identify the microscopic degree of freedom(s) responsible for the response, in contrast to the much cleaner spectral window between 70 and 90 µm. Nonetheless, the observed spectral weight of magnetization remains concentrated near the crystal-field resonance and is again comparatively weak at the positions of the strongest phonon modes. These observations suggest that the magnetic response near 30 µm predominantly originates from excitations with substantial crystal-field character, potentially enhanced through vibronic coupling with the lattice.

**Resonant quantum inverse Faraday effect**

Having identified crystal-field excitations as the dominant microscopic reservoir for optical angular-momentum transfer, we next ask whether they can also account for the observed spectral response. In particular, any microscopic description must explain both the resonant enhancement of THz-driven magnetization at the crystal-field transitions and its reversal of sign on opposite sides of the crystal-field resonance. To address this question, we construct a quantum theory of the inverse Faraday effect (see [25]-[26]) arising from resonant excitation between crystal-field levels of $Ce^{3+}$ [14]. The single 4*f* electron of $Ce^{3+}$ gives rise to a ground-state multiplet with total angular momentum $J = 5/2$. In $CeF_3$, the crystal field splits this multiplet into three Kramers doublets, denoted $g^{\pm}, e_1^{\pm}$, and $e_2^{\pm}$, as illustrated in Fig. 3A. Rather than corresponding to pure $|m_J >$ states, these doublets are crystal-field eigenstates composed of mixtures of the |±5/2>, |±3/2> and |±1/2> components. The crystal-field energies and wavefunctions are fixed by independent spectroscopic studies conducted almost half a century ago [14].

The quantum theory of the inverse Faraday effect gives the induced magnetic polarization as $M_{\mathrm{IFE}} = K(\omega)E^2$, where $M_{\mathrm{IFE}}$ is parallel to the wavevector of the THz electric field $E$ and $K(\omega)$ is the frequency-dependent inverse Faraday constant (see Supplementary Materials) [26]. The calculated $M_{\mathrm{IFE}}$ spectrum, shown in Fig. 3B, exhibits a pronounced dispersive feature centered on the $g^{\pm} \rightarrow e_1^{\mp}$ transition and a weaker resonance associated with the $g^{\pm} \rightarrow e_2^{\mp}$ transition, reproducing the dominant features of the experimental spectrum. Most remarkably, it predicts that the induced magnetization flips direction when detuned from the $g^{\pm} \rightarrow e_1^{\mp}$ resonance, in excellent agreement with the measurements. The close correspondence between experiment and theory therefore identifies resonantly excited crystal-field states as the microscopic origin of the induced magnetization.

Our calculations reveal that the microscopic origin of the sign reversal is rooted in the helicity-selective coupling between crystal-field states carrying opposite angular momentum. A net magnetization can be induced when the components of a Kramers doublet become unequally populated. According to the optical selection rules, circularly polarized THz light couples asymmetrically to the components of a Kramers doublet, exciting one component stronger than the other. For example, as illustrated in Fig. 3A, $\sigma^+$ excitation predominantly drives the $g^+ \rightarrow e_1^-$ transition, while the $g^- \rightarrow e_1^+$ transition is comparatively weaker. Thus, these excitations transfer angular momentum from the optical field to the crystal-field manifold by creating a nonequilibrium population imbalance, thereby generating a finite expectation value $\langle J_z \rangle$ and an associated magnetization. Assuming a crystal-field linewidth of 2.5 meV and a THz electric field of 0.13 MV/cm, the resulting magnetic polarization reaches ≈0.25 $\mu_B$ per $Ce^{3+}$ ion, corresponding to a substantial nonequilibrium polarization of the localized 4*f* moments. For comparison, generating an equivalent magnetization under equilibrium conditions would require an external magnetic field of about a tesla [27]. When the excitation wavelength is tuned through the resonance, the balance between the complementary helicity-allowed excitation pathways changes, reversing the population imbalance and consequently the sign of the magnetization.

The agreement between theory and experiment is particularly striking because the calculation includes only resonant crystal-field transitions and their associated inverse Faraday response, while entirely neglecting phononic and vibronic contributions [6],[12]. Nevertheless, our optical model reproduces both the spectral positions of the observed resonances and, most importantly, the reversal of the magnetization sign across the lowest crystal-field transition. The fact that such a minimal crystal-field model captures the dominant features of the experimental spectrum provides strong evidence that localized crystal-field excitations constitute the primary microscopic reservoir mediating optical angular-momentum transfer in $CeF_3$.

The spectral sign reversal is captured by the coherent resonant interaction underlying the inverse Faraday effect, yet the induced magnetization persists for up to ≈100 ps, substantially longer than the duration of the optical pulse. This behavior suggests that the optical excitation creates a long-lived population imbalance within the crystal-field manifold, allowing the transferred angular momentum to be stored after the coherent driving field has vanished. The subsequent relaxation likely occurs through nonradiative processes involving coupling between the localized 4f states and lattice degrees of freedom, consistent with the slow recovery of the equilibrium population distribution.

Additional interactions beyond the present theoretical description may further influence the magnetization dynamics and spectral response. Hybridization between the crystal-field excitation and a Raman-active optical phonon mode at 95 μm may contribute to the slower rise dynamics observed in Fig. 1 for the pumping wavelength of 88 μm (see Supplementary Materials) [12]. Changes in the optical response associated with an epsilon-near-zero region around 92 μm may further modify the local THz electric field experienced by the crystal-field excitations [28]-[30]. In contrast, the strong phonon absorption in the broad spectral range from 25 to 63 μm, enveloping the higher-energy crystal-field transition near 34 μm, may impede the underlying crystal-field contribution to the magnetization by introducing additional dissipation channels. While such effects are expected to modify the spectral weight and dynamics of the response, they do not alter the central conclusion that crystal-field excitations dominate the angular-momentum transfer process.

**Conclusions**

By mapping the helicity-dependent magneto-optical response of $CeF_3$ across a very broad range of the far-infrared spectrum, from 20 to 120 μm, we identify crystal-field excitations as a previously unrecognized route for dynamically converting optical angular momentum into magnetization. The observed sign reversal across the lowest crystal-field resonance provides a direct spectroscopic fingerprint of this mechanism and is reproduced by resonant electronic theory of the inverse Faraday effect. Our results establish localized crystal-field excitations as a microscopic reservoir for optical angular momentum and reveal a new pathway for generating magnetization with THz light. More broadly, they position localized electronic excitations as a platform for ultrafast optical control of magnetism and suggest that crystal-field engineering may provide a route toward designing THz-driven magnetic functionality in quantum materials.

## ACKNOWLEDGEMENTS

We thank all technical staff at FELIX for their technical support, Nils Dessmann and Dmytro Afanasiev for fruitful discussions, and Y.-L. Mathis and K. S. Rabinovich for support at the IR1 beamline of the Karlsruhe Research Accelerator (KARA). T.Z. and C.S.D. acknowledge support from the European Research Council ERC Grant Agreement No. 101115234 (HandShake), and A.K. acknowledges support from the European Research Council ERC Grant Agreement No.

101141740 (INTERPHON). M.S.M. and P.M.O. acknowledge support from the Knut and Alice Wallenberg Foundation (Grants No. 2022.0079 and No. 2023.0336) and from the EIC Pathfinder OPEN Grant No. 101129641 (OBELIX). The calculations were supported by the National Academic Infrastructure for Supercomputing in Sweden (NAISS) at NSC Linköping, funded by VR through Grant No. 2022-06725. The authors acknowledge the networking opportunities provided by the European COST Action No. CA23136 (CHIROMAG).

## AUTHOR CONTRIBUTIONS

C.S.D. conceived and managed the project. T.Z. built and performed the pump-probe experiments together with C.S.D. Y.L. and A.V.B. performed and processed the infrared ellipsometry measurements. M.S.M., M.W. and P.M.O. carried out the calculations. T.Z., A.K. and C.S.D. jointly discussed the results, and T.Z. and C.S.D. wrote the manuscript with contributions from all authors.

## COMPETING INTERESTS

The authors declare no competing interests.

## DATA AND MATERIALS AVAILABILITY

The data that support the findings of this study will be publicly available upon publication.

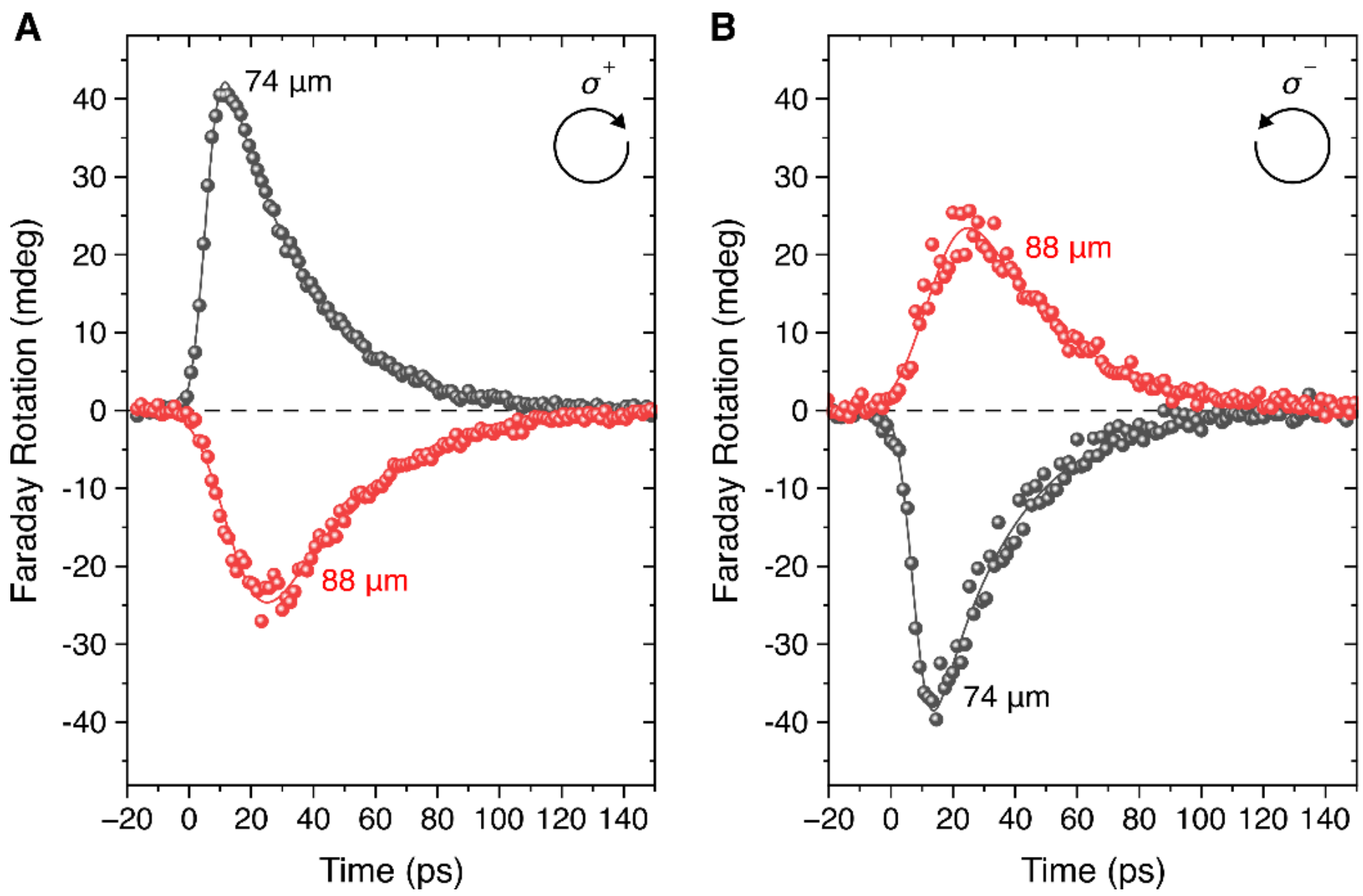

**Fig. 1 | Helicity- and wavelength-dependent ultrafast Faraday rotation in $CeF_3$.** Time-resolved pump-probe Faraday rotation measured following circularly polarized THz excitation at 74 and 88 μm for (**A**) left-handed ($\sigma^+$) and (**B**) right-handed ($\sigma^-$) optical helicity. Reversing the pump helicity inverts the transient response, whereas tuning the excitation wavelength across the lowest crystal-field resonance also reverses its sign even for fixed helicity.

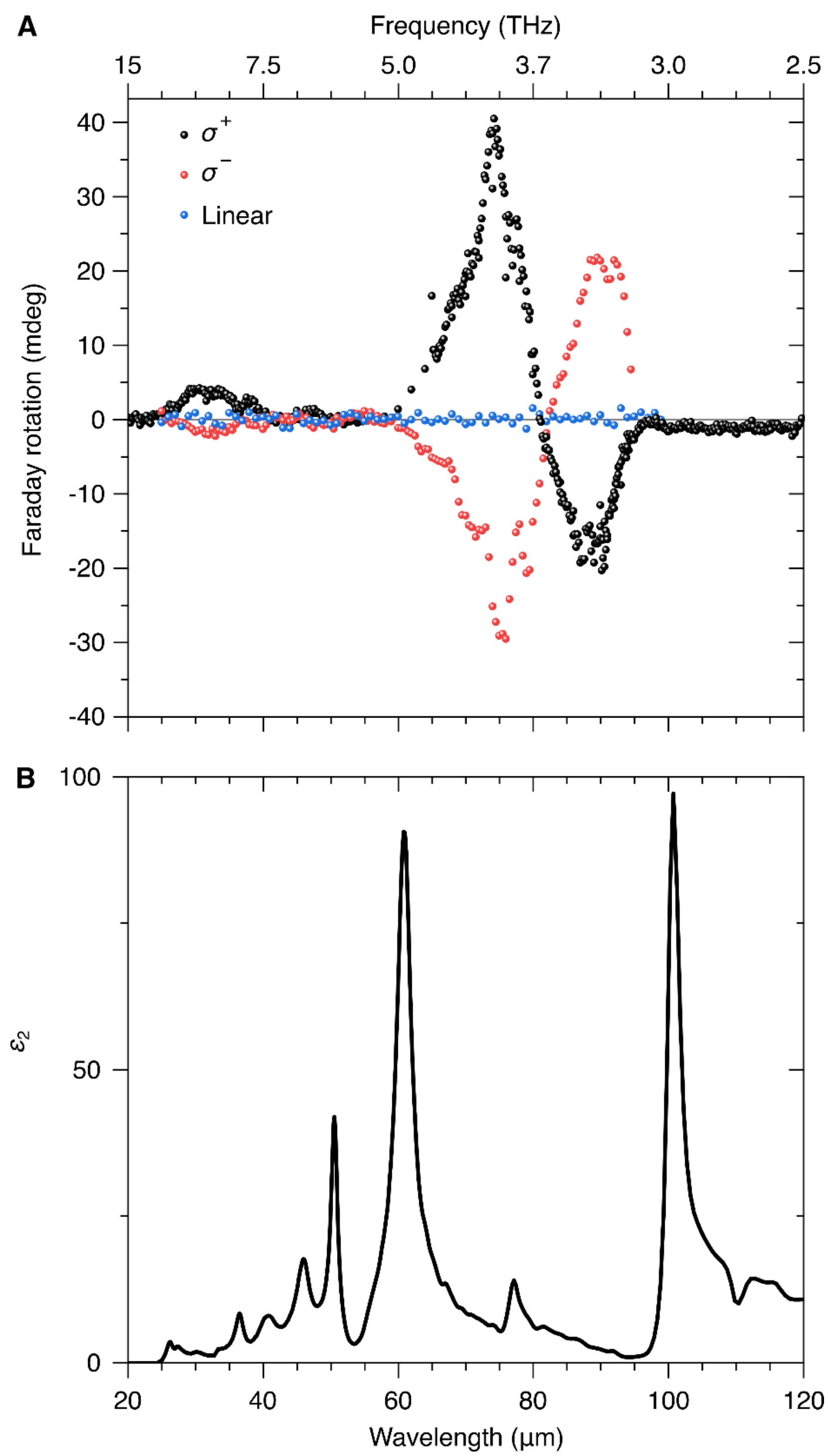

**Fig. 2| Spectral dependence of THz-driven magnetization and infrared optical excitations in $CeF_3$.** (**A**) Peak pump-induced Faraday rotation as a function of excitation wavelength for left-handed ($\sigma^+$) and right-handed ($\sigma^-$) circularly polarized excitation, and linearly polarized excitation. Reversing the optical helicity inverts the spectral response, whereas linearly-polarized excitation produces no measurable signal. (**B**) Imaginary part of the dielectric function $\varepsilon_2$ measured at 12 K by infrared ellipsometry. The strongest helicity-dependent magneto-optical response occurs near the lowest crystal-field excitation (71 μm) and does not scale with the doubly-degenerate optical phonon spectrum.

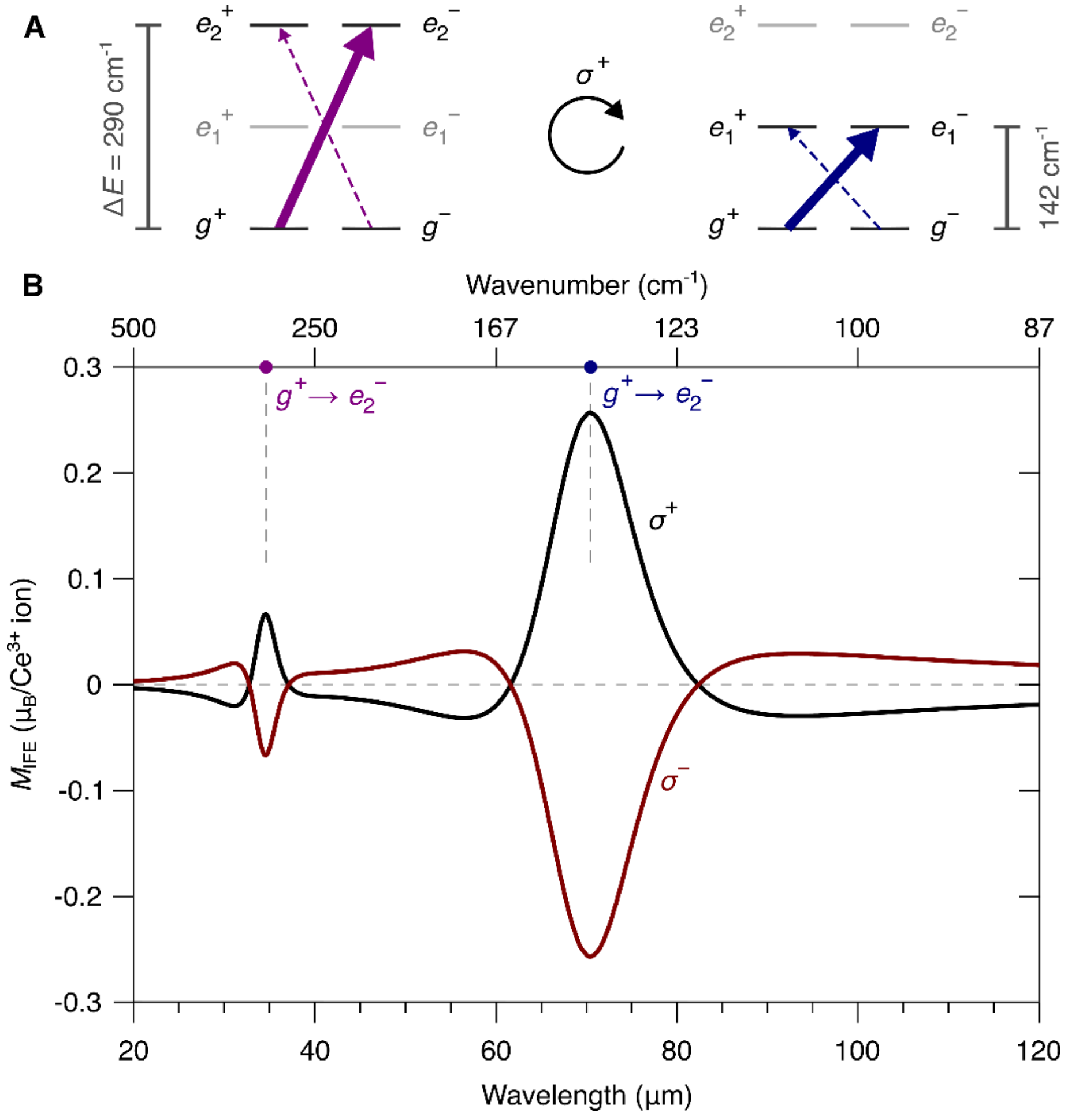

**Fig. 3 | Quantum inverse Faraday effect mediated by crystal-field excitations in CeF3. (A)** Schematic illustration of helicity-selective optical excitation between crystal-field states. Circularly polarized light dominantly excites one component of a Kramers doublets and generates a nonequilibrium population imbalance between the ground-state component $g^+$ and $g^-$, and the excited states $e_1^-$ and $e_1^+$, giving rise to a finite magnetization. (**B**) Calculated magnetization spectrum using a resonant electronic inverse Faraday effect. The model predicts a dominant dispersive resonance associated with the $g^\pm \rightarrow\ e_1^\mp$ crystal-field transition, including a reversal of the magnetization sign across the resonance, as well as a similar albeit weaker feature arising from the $g^\pm \rightarrow\ e_2^\mp$ transition.

# A crystal-field route to THz-driven magnetization

T. Zalewski,[1,2,*] M.S. Mrudul[3], Y. Lee[4], M. Weissenhofer[3], A.V. Boris[4], P.M. Oppeneer[3], A. Kirilyuk[1,2] and C.S. Davies[1,2,*]

[1] *HFML-FELIX, Radboud University, Toernooiveld 7, 6525 ED Nijmegen, The Netherlands*

[2] *Radboud University, Institute for Molecules and Materials, 135 Heyendaalseweg, 6525 AJ Nijmegen, The Netherlands*

[3] *Department of Physics and Astronomy, Uppsala University, P.O. Box 516, SE-75120 Uppsala, Sweden*

[4] *Max Planck Institute for Solid State Research, Heisenbergstrasse 1, 70569 Stuttgart, Germany*

* Corresponding authors: tomasz.zalewski@ru.nl and carl.davies@ru.nl

## SUPPLEMENTARY MATERIALS

### MATERIALS AND METHODS

1. Materials

Single-crystalline $CeF_3$ ($5 \times 5 \times 0.5$ mm$^3$) was purchased commercially from MSE Supplies. The crystal was double-side polished with a (001) crystal orientation, such that the crystallographic c-axis (optical axis) was normal to the surface. Measurements were also repeated on thicker crystals of $CeF_3$, with thickness 2 mm, purchased commercially from Advatech. The results obtained from both crystals were indistinguishable.

2. Two-color pump-probe setup

Far-infrared pump pulses were generated using the FEL-1 beamline of the free-electron lasers at HFML-FELIX (Nijmegen, The Netherlands) [S1]. FEL-1 provides wavelength-tunable radiation in the spectral range 25-120 μm with an experimentally adjustable bandwidth of 0.35-5%. The spectral distribution of the infrared pulses was monitored continuously during the experiments using an inline spectrometer that measured a fraction of the FEL radiation. Figure S1 shows a characteristic spectrum of the pump pulses used to obtain the measurements presented in Fig. 1. The pulses were delivered in the form of 10-μs-long macropulses at a repetition rate of 10 Hz, consisting of a train of micropulses separated by 40 ns, i.e., at 25 MHz repetition rate.

The FEL radiation is intrinsically linearly polarized. Circular polarization was generated by first rotating the linear polarization to ±45° using an image rotator [S1], and subsequently transmitting the beam through a right-angle silicon prism acting as a broadband phase retarder. Under total internal reflection, the p- and s-polarized components acquire a relative phase shift close to $\pi/2$, resulting in nearly circularly polarized output radiation [S2]. The polarization state was characterized using an analyzer and a liquid-nitrogen-cooled HgCdTe detector, with Figure S2 showing the normalized transmitted intensity as a function of analyzer angle. The transmitted intensity varies between 72% and 100% of its maximum value, leading to the pulse having elliptical polarization with a minor-to-major electric-field amplitude ratio of $\sqrt{0.72}$ = 0.85. Owing to the negligible wavelength dependence of the refractive index of silicon in the far-infrared regime, the pump helicity is preserved over the entire tuning range of the free-electron laser. This is vital for measuring the spectrum shown in Fig. 2 in the main text, since the helicity must be fixed with certainty while scanning the wavelength of the far-infrared radiation from 20 to 120 μm.

The circularly polarized far-infrared pulses were focused onto the sample using an off-axis parabolic mirror with focal-length 50.8 mm, resulting in an approximately Gaussian spot with a diameter of ≈60 μm at a wavelength of 12 μm, as determined by the Liu method [S3]. The spot size is assumed to scale linearly with wavelength. The sample was mounted in a flow cryostat equipped with polyethylene windows, which provide broad transmission in the far-infrared range while introducing negligible static and immeasurable dynamic birefringence.

The far-infrared pulses are transported from FEL-1 to the sample through a vacuum-pumped optical beamline and a nitrogen-purged pump-probe setup. A relative humidity level of less than 2% was measured next to the sample position for all results shown here.

Transient magnetization dynamics were monitored using synchronized visible probe pulses derived from an amplified oscillator (MENLO Orange HP). This laser produced pulses of approximately 150 fs duration centered at 1040 nm, which were frequency doubled by second-harmonic generation to obtain probe pulses at 520 nm. The oscillator repetition rate of 100 MHz was synchronized to the FEL micropulse train. A retroreflector mounted upon a motorized delay stage was used to obtain temporal resolution.

The linearly polarized probe beam was focused by a lens (focal length 100 mm) onto the sample via a through-hole in the parabolic mirror and spatially overlapped with the THz pump pulse at normal incidence on the $CeF_3$ crystal, with both beams propagating collinearly. After transmitting through the crystal and rear cryostat window, the probe beam was recollimated, passed through a zero-order half-waveplate for polarization balancing, and separated into its orthogonal polarization components by a polarizing beam splitter. The two components were then detected by a balanced pair of high-bandwidth photodetectors (HBPR-500M-10K-SI-FST, FEMTO), yielding a difference signal proportional to pump-induced polarization rotations. Such helicity-dependent Faraday rotations have previously been established as a signature of transient magnetization in $CeF_3$ [S4]. Contributions from non-magnetic pump-induced birefringence are excluded by the helicity-odd response, absence of measurable signals for linearly polarized excitation, and resonant enhancement at crystal-field transitions carrying electronic angular momentum. A schematic of the experimental arrangement is shown in Fig. S3.

To improve the signal-to-noise ratio further, the detector output was demodulated using a high-frequency lock-in amplifier (UHFLI, Zurich Instruments) referenced to the micropulse repetition rate of 25 MHz. Gated acquisition was employed such that only the signal collected during the duration of the macropulse contributed to the measurement.

3. <u>Spectroscopic ellipsometry</u>

The low-temperature optical response of $CeF_3$ (measured at 12 K) was determined using synchrotron-based spectroscopic ellipsometry [S5]. Measurements in the frequency range 60 to 700 $cm^{-1}$ (14-167 μm) were performed using homemade ellipsometers in combination with a Bruker IFS 66v/S Fourier-transform infrared spectrometer. To obtain infrared radiation in this spectral range, we utilized synchrotron edge radiation from the 2.5-GeV electron storage ring at the IR1 beamline of the Karlsruhe Research Accelerator at the Karlsruhe Institute of Technology (Germany). The ellipsometric parameters $\Psi$ and $\Delta$ were measured at an angle of incidence of 75° relative to the normal to the *ab* or (*a*/*b*)*c* plane. These quantities define the complex reflectance ratio $r_p/r_s = \tan(\Psi)\exp(i\Delta)$, where $r_p$ and $r_s$ denote the complex Fresnel reflection coefficients for light polarized parallel and perpendicular to the plane of incidence, respectively. Sample orientations were selected with the $p$ component of the electric field vector parallel to the *a* (or *b*) axis.

The real and imaginary parts of the complex dielectric function $\varepsilon(\omega) = \varepsilon_1(\omega) + i\varepsilon_2(\omega)$, presented in Fig. S4, were obtained by direct inversion of $\Psi$ and $\Delta$ assuming semi-infinite bulk behavior of the $CeF_3$ crystal. The pseudo-dielectric function provides the *a*- (or *b*-) axis tensor element of the dielectric tensor $\varepsilon^a = \varepsilon^b$ with a minor contribution of the orthogonal components of the dielectric tensor $\varepsilon^c$. The infrared-active transverse optical (TO) phonons appear as resonances in $\varepsilon_2(\omega)$, while the longitudinal optical phonons are identified by peaks in the energy-loss function Im($-\varepsilon^{-1}$). The observed in-plane phonon modes are listed in Table S1. The absorption coefficient was calculated using $\alpha(\omega) = 4\pi k/\lambda$, where $k$ is the extinction coefficient obtained from the complex refractive index $\hat{n} = n + \mathrm{i}k = \sqrt{\varepsilon(\omega)}$. Previous spectroscopic studies have established that the observed eleven TO phonon modes possess doubly-degenerate ($E_\mathrm{u}$) symmetry [S6].

4. Crystal field energies and wavefunctions of cerium trifluoride

The crystal-field level scheme of $CeF_3$ has been studied for more than half a century. These studies consistently show that the $Ce^{3+}$ ion, with a single 4*f* electron, possesses a $^2F_{5/2}$ ground-state multiplet that is split by the crystal field of the fluorine ions into three Kramers doublets [S7]. The resulting crystal-field transitions, positioned in the far-infrared spectral range, form the basis of the model presented in the main text.

The energies and wavefunctions employed in our calculations are taken directly from the crystal-field literature and are not treated as adjustable parameters. In the $|J = 5/2, m_J >$ basis, the crystal-field eigenstates correspond to linear combinations of the $m_\mathrm{J} = \pm 1/2$, $\pm 3/2$, and $\pm 5/2$ components. The energies and wavefunctions used throughout this work are reproduced in Table S2 from Ref. [S8].

5. Quantum inverse Faraday model

A quantum-mechanical description of the inverse Faraday effect based on density-matrix perturbation theory, second order in the electric field, was developed in Ref. [S9]. It has been previously applied to calculate the coherent laser-induced magnetization in metals [S10]-[S11]. The induced magnetization can be written as $M_{IFE} = M_{o,z} + M_{dA,z} + M_{dB,z} = (K_{o,z} + K_{dA,z} + K_{dB,z} + c.c.)E^2$, where $E$ is the THz electric field amplitude, and the inverse Faraday coefficients $(K_{o,z}, K_{dA,z}, K_{dB,z})$ can be expressed in matrix-element notation as follows:

$$K_{o,z} = e^2 \sum_{m \neq n;l} M_{mn,z} \frac{\dfrac{r_{nl}^{+} r_{lm}^{-} (\rho_{mm}^0 - \rho_{ll}^0)}{\varepsilon_l - \varepsilon_m + i\hbar\Gamma_{lm} - \hbar\omega} - \dfrac{r_{nl}^{-} r_{lm}^{+} (\rho_{ll}^0 - \rho_{nn}^0)}{\varepsilon_n - \varepsilon_l + i\hbar\Gamma_{nl} - \hbar\omega}}{\varepsilon_n - \varepsilon_m + i\hbar\Gamma_{nm}}, \tag{S1}$$

$$K_{dA,z} = e^2 \sum_{n,l} M_{nn,z} \left( \frac{r_{nl}^{+} r_{ln}^{-} (\rho_{ll}^0 - \rho_{nn}^0)}{(\varepsilon_l - \varepsilon_n + i\hbar\Gamma_{ln} - \hbar\omega)^2} + \frac{r_{nl}^{-} r_{ln}^{+} (\rho_{ll}^0 - \rho_{nn}^0)}{(\varepsilon_n - \varepsilon_l + i\hbar\Gamma_{nl} - \hbar\omega)^2} \right), \tag{S2}$$

$$K_{dB,z} = e^2 \sum_{n,l} \frac{M_{nn,z} r_{nl}^{+} r_{ln}^{+} (\rho_{nn}^0 - \rho_{ll}^0)(i\hbar\Gamma_{ln} - \hbar\omega)}{\hbar\omega[(\varepsilon_l - \varepsilon_n)^2 + (\hbar\Gamma_{ln} + i\hbar\omega)^2]}. \tag{S3}$$

Here $\hbar\omega$ is the energy of the incident laser field and $\varepsilon_n$ denotes the energy of state $|n\rangle$. The matrix elements of the magnetization operator are defined as $M_{mn,z} = \langle m|\hat{M}_z|n\rangle$. The $\rho_{nm}^0$ denote the matrix elements of the equilibrium density matrix. The incident light is assumed to propagate

along the $z$-axis, such that the relevant optical transitions are governed by the circular components of the electric dipole operator, $-e\hat{r}^{\pm} = -e(\hat{x} \pm i\hat{y})/\sqrt{2}$, with the corresponding matrix elements $-er^{\pm}_{nl}$. The quantity $\hbar\Gamma_{ln}$ is a phenomenological broadening parameter introduced to account for dephasing.

As discussed in the previous subsection, the crystal field states of $Ce^{3+}$ can be expressed as linear combinations of the angular momentum basis states $|J, m_J\rangle$, with $J = 5/2$ (Ref. [S8]). The matrix elements between these angular momentum states are evaluated using the Wigner-Eckart theorem. For the crystal field states of $Ce^{3+}$, the radial part of the $4f$-electron wavefunctions is approximated using the Hartree-Fock wavefunctions reported in Ref. [S12]. In the present work, we assume a state-independent, constant broadening $\hbar\Gamma$ of 2.5 meV (see Fig. S5 for the dependence of our results on the broadening). The peak electric field is set to 0.13 MV/cm, consistent with the experiment.

## SUPPLEMENTARY TEXT

### 1. Discussion of dependences for fixed sample temperature, and method of fitting

In the pump-probe measurements, we additionally studied how the measured signals depend on the energy of the THz excitation. In Fig. S6A-B, we show time-resolved traces obtained with circularly polarized pump pulses centered at 74 and 88 µm, respectively. To quantify how the strength and decay time of the measured signals depend on the pulse energy, we fit the observed time-resolved signals $\Delta\theta_{signal}$ with the convolution of a Gaussian pulse with an exponential decay, given by

$$\Delta\theta_{signal} = \frac{A}{2}\exp\left[\frac{\sigma^2}{2\tau^2} - \frac{t-\mu}{\tau}\right]\mathrm{erfc}\left[\frac{\sigma^2 - \tau(t-\mu)}{\sigma\tau\sqrt{2}}\right], \quad \text{(S4)}$$

where erfc is the complementary error function. Here, $A$ is the amplitude, $\tau$ is the lifetime of the exponential component, and $\sigma^2$ and $\mu$ is the variance and mean of the Gaussian component, respectively. This fitting equation captures well the behavior observed in all time-resolved traces.

In Fig. S6C, we present the fitted amplitude of the recorded signals as a function of the pulse energy. We observe that, for both pump wavelengths, the signal amplitude grows linearly with pump energy. Since the pulse energy is proportional to the square of the electric field, we therefore conclude that the observed signal scales quadratically with the peak electric field. To calculate the latter, we assume the THz pulse has a duration of 5 ps and is focused to a spot of diameter ≈400 µm.

In Fig. S6D, we also present the energy-dependence of the fitted decay rate. We observe that the decay rate, for both pump wavelengths, is broadly independent of the pump energy. Together with the linear increase of the signal amplitude, this indicates that the measurements are performed within a linear regime, consistent with optical pumping of crystal-field excitations rather than a fluence-dependent thermal or nonlinear process.

2. Phononic and vibronic excitations in cerium trifluoride

The low-energy excitation spectrum of $CeF_3$ has been extensively investigated using Raman [S9], infrared [S6], and neutron spectroscopy [S14]. A central finding of these studies is that the crystal-field excitations of the $Ce^{3+}$ 4*f* manifold are strongly coupled to lattice vibrations, giving rise to pronounced magnetoelastic and vibronic effects.

Early Raman measurements by Schaack and co-workers revealed that doubly degenerate $E_g$ phonons split in an applied magnetic field [S15]. The splitting was interpreted as a magnetoelastic self-energy effect arising from coupling between the lattice vibrations and the crystal-field-split 4*f* electronic states of $Ce^{3+}$ [S7]-[S8]. In this picture, the two circular components of an $E_g$ phonon interact differently with the magnetic-field-polarized 4*f* multiplet, producing a phonon Zeeman splitting [S16]-[S18]. These experiments provided some of the earliest evidence that the lattice dynamics of rare-earth fluorides are strongly influenced by the underlying crystal-field degrees of freedom.

Subsequent Raman studies established that several crystal-field excitations and Raman-active phonons occur in close spectral proximity [S7]-[S8]. In particular, the crystal-field excitation near 34 µm lies within a dense population of Raman-active phonons, leading to hybridization between vibrational and electronic degrees of freedom. As a result, the eigenmodes in this spectral region are more appropriately described as vibronic excitations rather than as purely phononic or purely electronic modes.

To assess the possible influence of these vibronic effects on the light-induced magnetization, Fig. S7 compares the measured magneto-optical spectrum with that of Raman-active phonons, of different symmetries, previously reported in the literature. The strongest magneto-optical response in the spectral range from 20 to 40 µm occurs in the vicinity of the 34 µm crystal-field excitation, where substantial crystal-field-phonon hybridization is expected. Consequently, the microscopic origin of the response in this region cannot be uniquely assigned to either crystal-field or phononic degrees of freedom. Instead, the observed magnetization likely reflects the mixed vibronic character of the underlying excitations.

This situation contrasts with the lower-energy crystal-field transition at 71 μm. In the spectral window between 70 and 90 μm, the magneto-optical response exhibits a pronounced sign reversal across an isolated crystal-field resonance, occurring in a region that is devoid of strong Raman-active phonons. The spectral assignment here is therefore considerably less ambiguous, providing the strongest evidence that crystal-field excitations constitute the dominant microscopic reservoir mediating optical angular-momentum transfer in $CeF_3$.

3. Spectral comparison of magnetization and infrared-active phonons near 30 μm

To further examine the microscopic origin of the helicity-dependent magnetization in the spectral region previously associated with circularly polarized phonons in Ref. [S4], Fig. S8 shows a magnified comparison between the measured magneto-optical response and the imaginary part of the dielectric function $\varepsilon_2$ between 20 and 40 μm. This spectral range contains several strongly absorbing infrared-active phonon modes as well as the higher-energy crystal-field excitation of $Ce^{3+}$ near 34 μm. Note that the measurements presented here were obtained by fits of the time-delayed transients recorded for each excitation wavelength, with fixed helicity.

Although a helicity-dependent magnetic response is observed in this wavelength range, its spectral dependence does not appear to be governed by the oscillator strengths of the infrared-active phonons. The largest magnetization occurs in the vicinity of the crystal-field excitation near 34 μm, whereas several strongly absorbing phonon resonances produce only weak magneto-optical responses. This absence of a direct correlation between phonon spectral weight and THz-driven magnetization indicates that the transfer of angular momentum cannot be understood solely from the excitation of infrared-active phonons. However, the high density of infrared-active phonons, Raman-active modes, and crystal-field excitations in this spectral region prevents a complete separation of electronic and lattice contributions. The observed response is therefore most consistent with excitations possessing substantial crystal-field character, potentially enhanced through vibronic coupling between the $Ce^{3+}$ 4*f* states and the surrounding lattice.

4. Physical mechanism of the light-induced magnetization

We present the separate contributions of the $K_{o,z}$, $K_{dA,z}$, and $K_{dB,z}$ terms [defined in Eqs. (S1)-(S3)] to the light-induced magnetization in Fig. S9. The results clearly show that the experimentally observed features are well explained by the $K_{dA,z}$ term. Under appropriate assumptions, Eq. (S2) can be written as,

$$K_{dA,z} = e^2 \sum_{n,l\ (\varepsilon_n > \varepsilon_l)} \left(M_{nn,z} - M_{ll,z}\right) \frac{|r_{nl}^-|^2 (\rho_{ll}^0 - \rho_{nn}^0)}{(\varepsilon_n - \varepsilon_l + i\hbar\Gamma_{nl} - \hbar\omega)^2} . \quad \text{(S5)}$$

The electric dipole transition matrix elements between angular momentum states satisfy $\langle J, m_f | r^{\pm} | J, m_i \rangle \propto \delta_{m_f, m_i \pm 1}$. Consequently, for circularly polarized light, electric dipole transitions between $Ce^{3+}$ crystal field states obey the following selection rules: transitions to the time-reversal-alike excited states ($|g^{\pm}\rangle \rightarrow |e_1^{\pm}\rangle$ and $|g^{\pm}\rangle \rightarrow |e_2^{\pm}\rangle$) are forbidden, whereas transitions to the time-reversal-conjugate excited states ($|g^{\pm}\rangle \rightarrow |e_1^{\mp}\rangle$ and $|g^{\pm}\rangle \rightarrow |e_2^{\mp}\rangle$) are allowed. We can write the magnetization induced by the $|g^{\pm}\rangle \rightarrow |e_1^{\mp}\rangle$ optical transitions as,

$$M_z^{g \rightarrow e_1} = e^2 E^2 \frac{\Delta M_{eg}}{(\Delta\varepsilon_{eg} + i\hbar\Gamma_{nl} - \hbar\omega)^2} \left( |\langle e_1^+ | \hat{r}^+ | g^- \rangle|^2 - |\langle e_1^- | \hat{r}^+ | g^+ \rangle|^2 \right), \quad \text{(S6)}$$

where $\Delta M_{eg} = \langle e_1^+ | \widehat{M}_z | e_1^+ \rangle - \langle g^- | \widehat{M}_z | g^- \rangle$, and $\Delta\varepsilon_{eg} = \varepsilon_{e_1} - \varepsilon_g$.

This equation shows that $\sigma^+$ polarized excitation allows electric dipole transitions $|g^+\rangle \rightarrow |e_1^-\rangle$ and $|g^-\rangle \rightarrow |e_1^+\rangle$, whereas each of these transitions induces an opposite change in magnetization. Since the dipole matrix elements $\langle e_1^+ | \hat{r}^+ | g^- \rangle$ and $\langle e_1^- | \hat{r}^+ | g^+ \rangle$ are different in magnitude, a population imbalance is generated, resulting in a net magnetization. When the helicity of the laser pulse is reversed to $\sigma^-$, the above expression acquires an overall negative sign since the transition amplitudes are interchanged, i.e., $|\langle e_1^- | \hat{r}^- | g^+ \rangle|^2 = |\langle e_1^+ | \hat{r}^+ | g^- \rangle|^2$, and $|\langle e_1^+ | \hat{r}^- | g^- \rangle|^2 = |\langle e_1^- | \hat{r}^+ | g^+ \rangle|^2$. The same arguments hold for excitations to the second excited state $e_2^{\pm}$. An illustration of this mechanism is presented in Fig. S10. The differences in the strength of induced magnetization observed near the first and the second excited states arise from differences in the $|m_J\rangle$ composition of the excited states as well as differences in the dipole coupling.

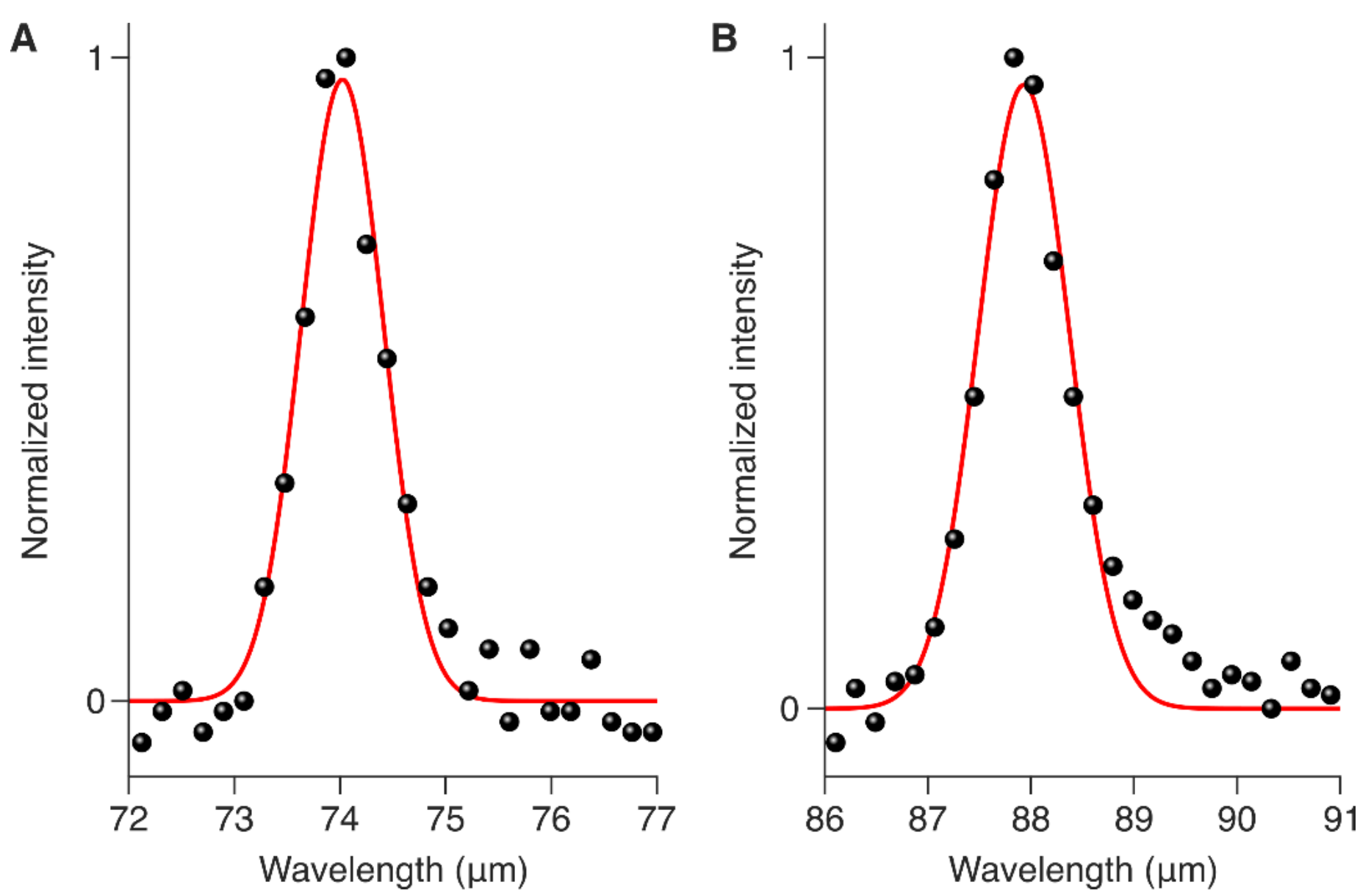

**Supplementary Fig. S1 | Representative spectral characterization of the far-infrared pump pulses.** Normalized spectra of the FELIX pump pulses used while measuring the time-resolved measurements shown in Fig. 1, at center wavelengths **(A)** 74 µm and **(B)** 88 µm. Black symbols show the measured spectral intensity, and red curves show Gaussian fits used to estimate the pump bandwidth. The fitted rms spectral bandwidths are 1.1% and 1.0% of the center wavelength for the 74 µm and 88 µm pump pulses, respectively, corresponding to less than 900 nm.

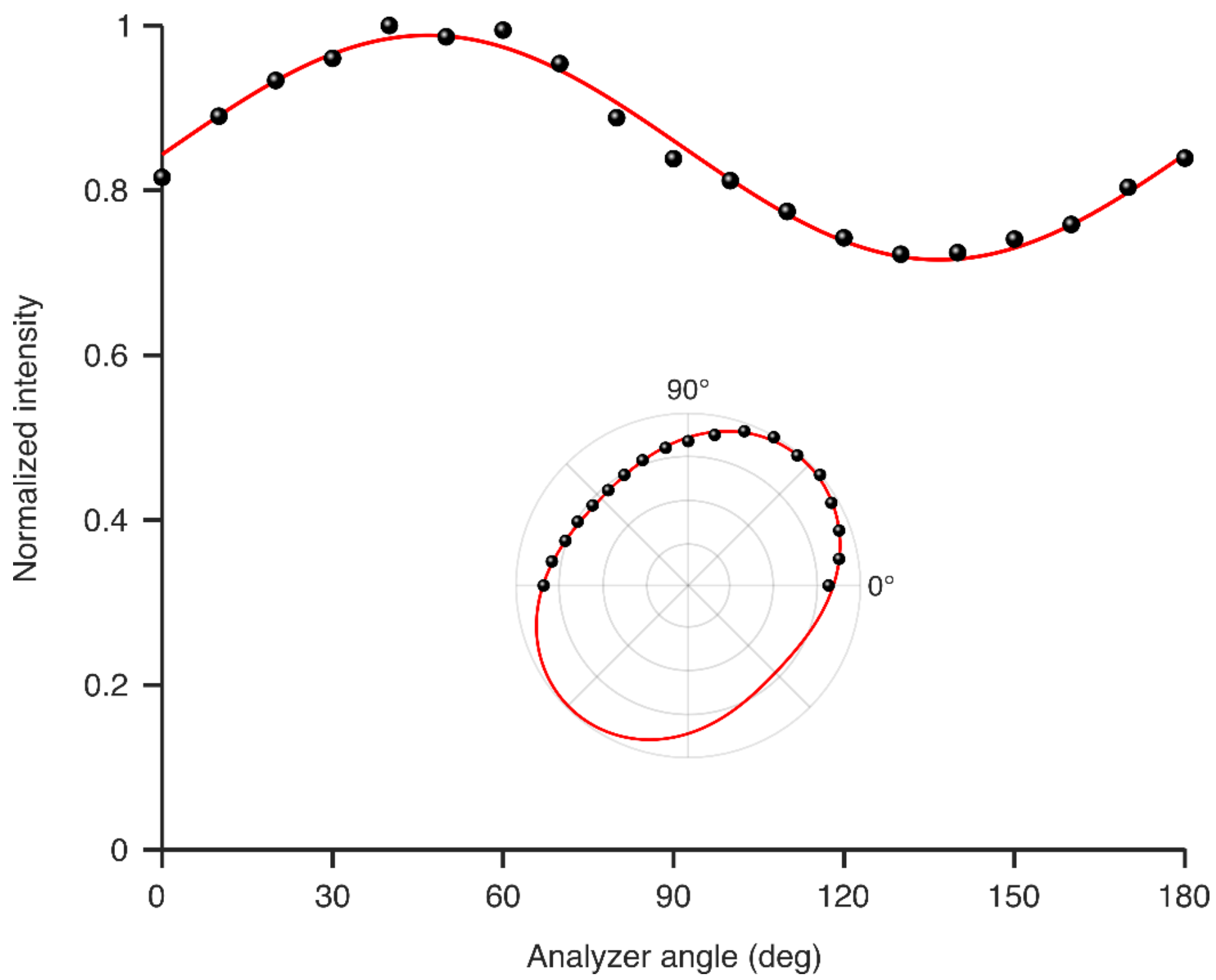

**Supplementary Fig. S2 | Characterization of the far-infrared pump polarization generated by the silicon phase retarder.** Normalized transmitted intensity at 33 μm measured as a function of analyzer angle using a rotating wire-grid polarizer and a $N_2$-cooled HgCdTe detector. The red curve shows a sinusoidal fit to the experimental data (black circles). Inset: Polar representation of the analyzer scan, with measured data shown for analyzer angles between 0° and 180° and the fitted response extended over the full 360°.

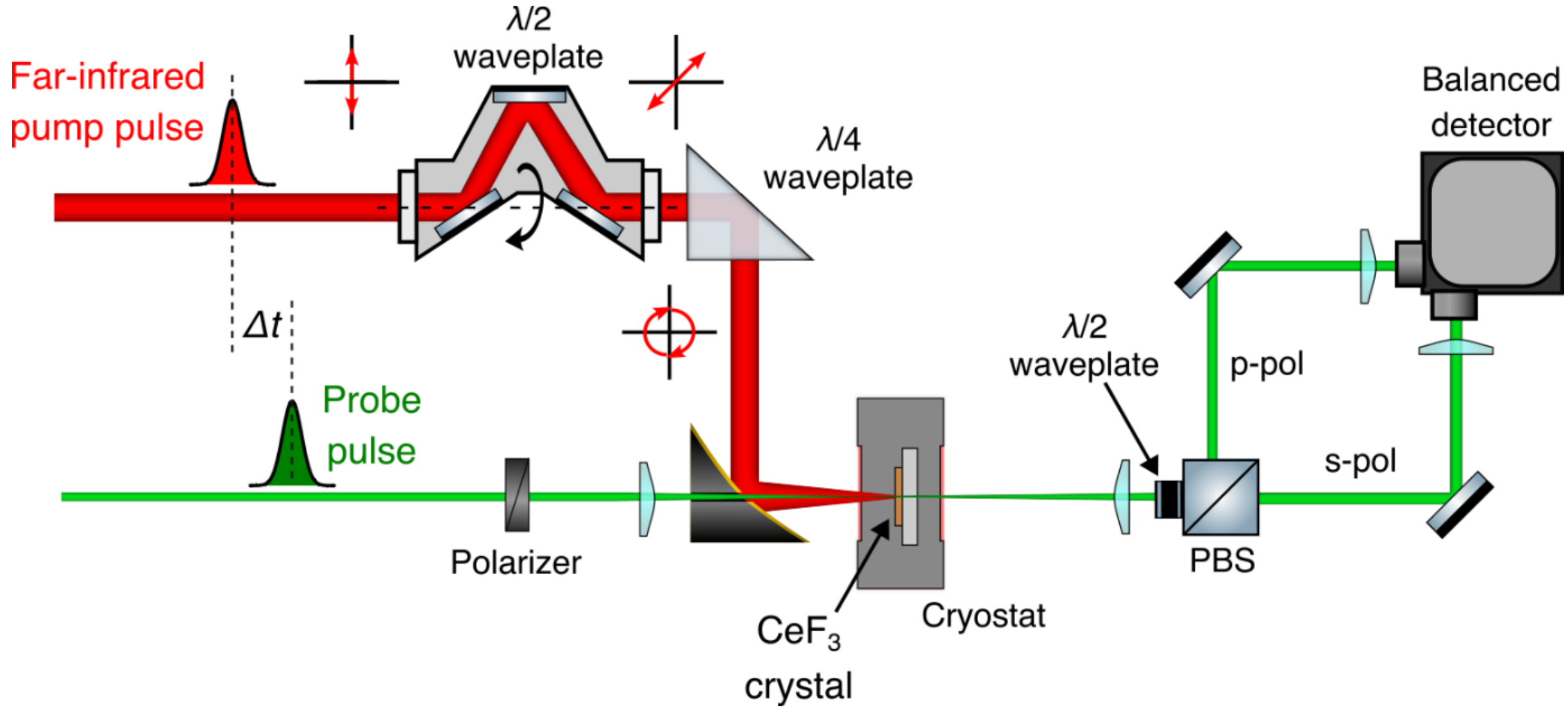

**Supplementary Fig. S3 | Experimental pump-probe setup.** Schematic of the wavelength-selective pump-probe Faraday spectroscopy experiment. As indicated by the polarization insets, the linearly polarized far-infrared pump pulses ($20 < \lambda < 120$ μm) delivered by the free-electron laser are converted to circular polarization by a combination of an image rotator and right-angle silicon prism, being equivalent to half-wave and quarter-wave waveplates, respectively, before being focused onto the $CeF_3$ crystal. A time-delayed probe pulse (520 nm, ≈ 150 fs) passes a polarizer and is focused on the sample. After transmitting through the $CeF_3$ crystal, the probe pulse is separated into orthogonal polarization components by a polarizing beam-splitter (PBS). A balanced detector measures the differential signal between the two polarization channels, yielding the pump-induced Faraday rotation.

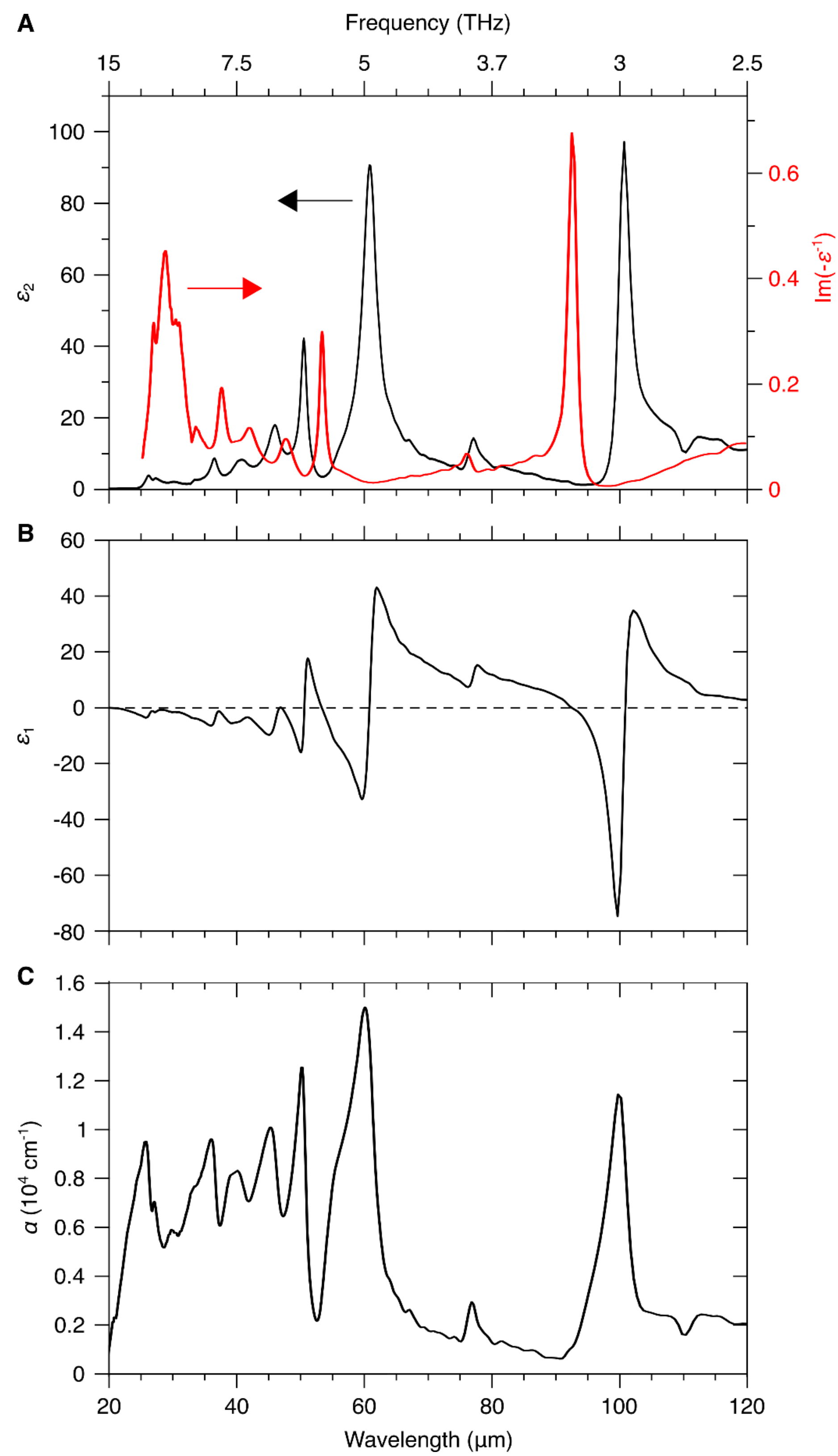

**Supplementary Fig. S4 | Optical properties of $CeF_3$ measured by infrared ellipsometry. (A)** Imaginary part of the dielectric function $\varepsilon_2$ (black) and loss function Im(-1/$\varepsilon$) (red). **(B)** Real part of the dielectric function $\varepsilon_1$. **(C)** Absorption coefficient $\alpha$ calculated from the dielectric function. The optical response was measured with the electric field polarized perpendicular to the crystallographic c-axis at 12 K.

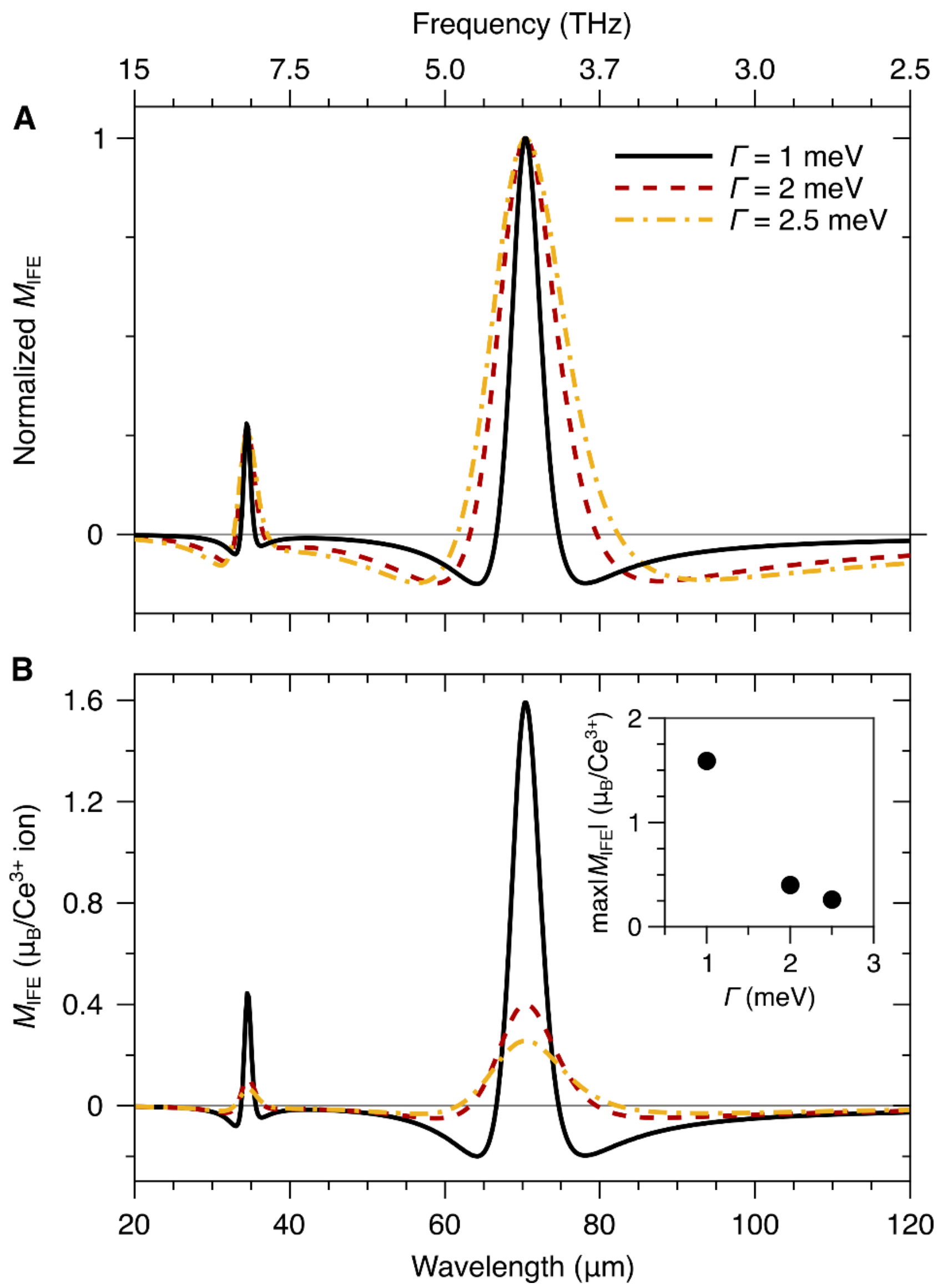

**Supplementary Fig. S5 | Robustness of the calculated inverse Faraday magnetization against the assumed linewidth. (A)** Calculated $\sigma^+$ spectra for varying linewidths $\Gamma$ as indicated, normalized to their respective maximum absolute values. The linewidth primarily broadens the resonances while preserving the overall dispersive spectral dependence and wavelength-dependent sign reversal. **(B)** The corresponding spectra on an absolute scale, showing that increasing the linewidth suppresses the magnitude of the resonantly enhanced magnetization without qualitatively altering the spectral response. Inset: Maximum calculated inverse Faraday magnetization as a function of $\Gamma$, illustrating the reduction in peak magnetization with increasing linewidth.

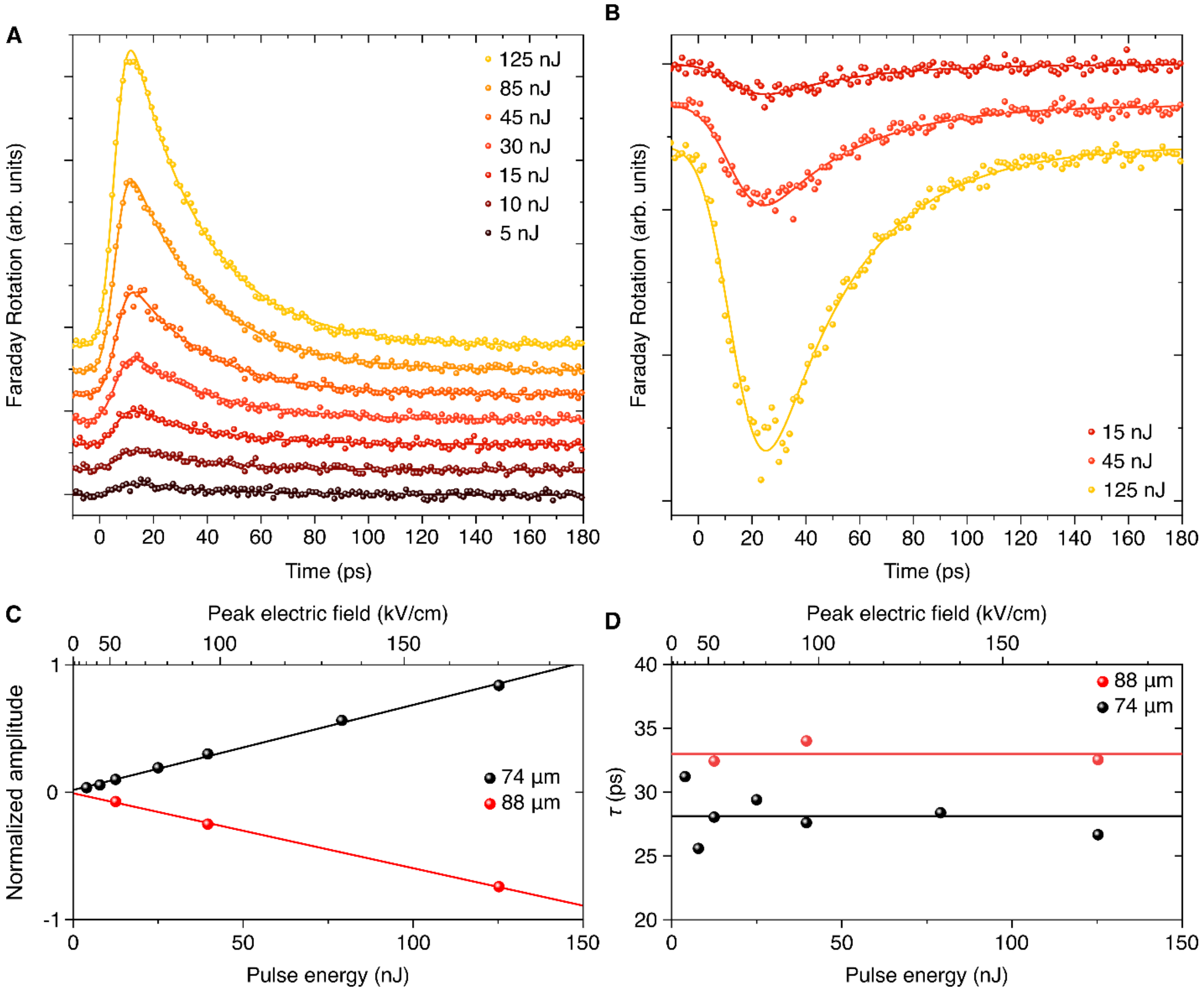

**Supplementary Fig. S6 | Pulse-energy-dependence of the light-induced magnetization. (A)-(B)** Time-resolved Faraday rotation measured at 4 K following excitation with a circularly polarized far-infrared pulse with varying energy as indicated, at 74 and 88 µm, respectively. The signals are offset relative to each other, and solid lines are fits to the experimental data according to Eq. (S4). **(C)-(D)** The fitted amplitude and characteristic decay constant of the transient Faraday rotation as a function of pulse energy, respectively, for 74 and 88 µm as indicated. The lines are linear fits.

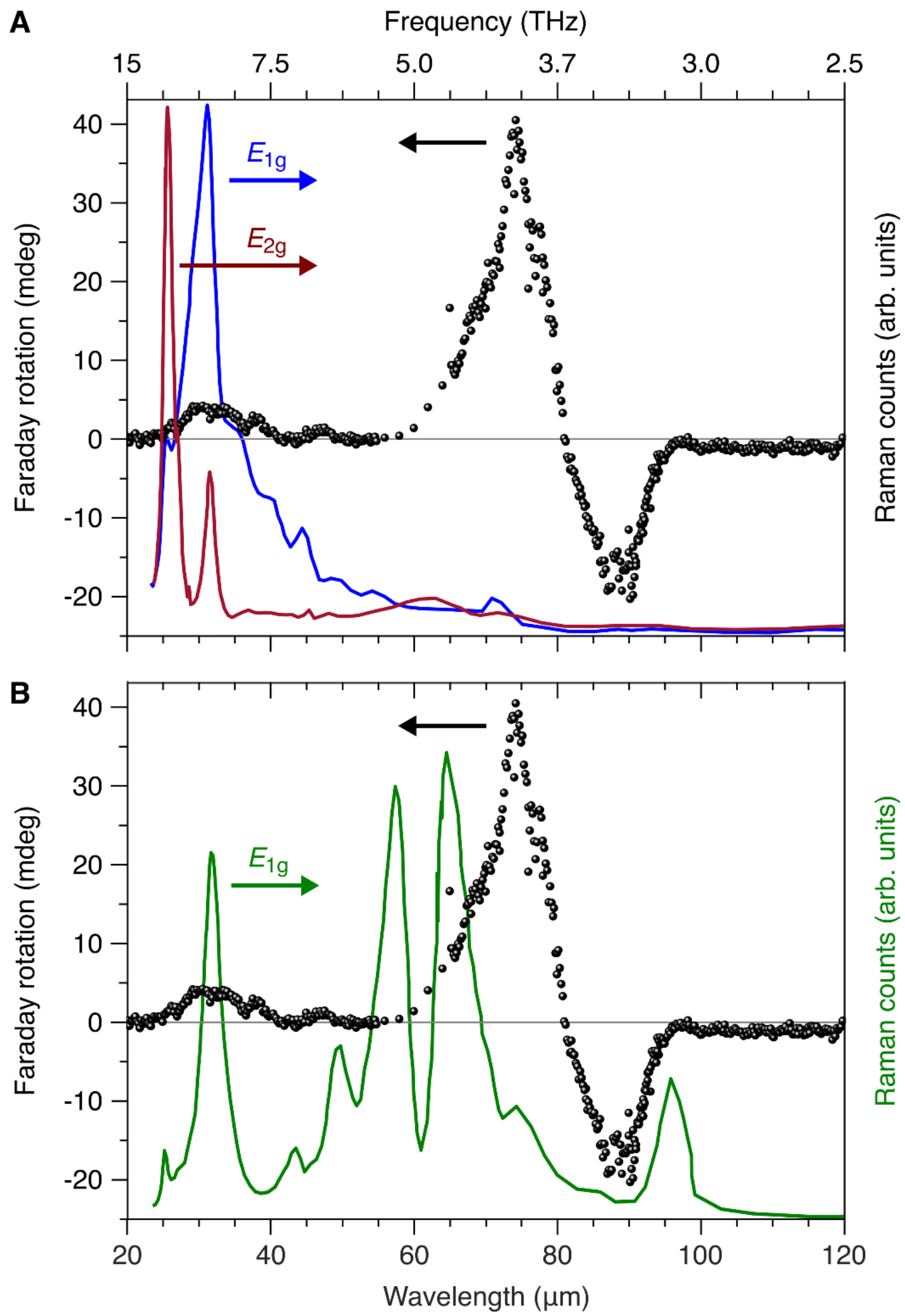

**Supplementary Fig. S7 | Comparison of the THz-driven magnetization spectrum with previously-reported Raman spectra of $CeF_3$.** The Faraday rotation spectrum measured in this work (black symbols) is compared with Raman spectra reported in the literature for different polarization geometries. **(A)** Raman spectrum measured in the X(YX)Y and X(YZ)Y scattering configurations, corresponding to $E_{2g}$ and $E_{1g}$ symmetry, respectively. **(B)** Raman spectrum measured in the X(ZX)Y scattering configuration, corresponding to $E_{1g}$ symmetry. The Raman spectra were digitized from Refs. [S7]-[S8].

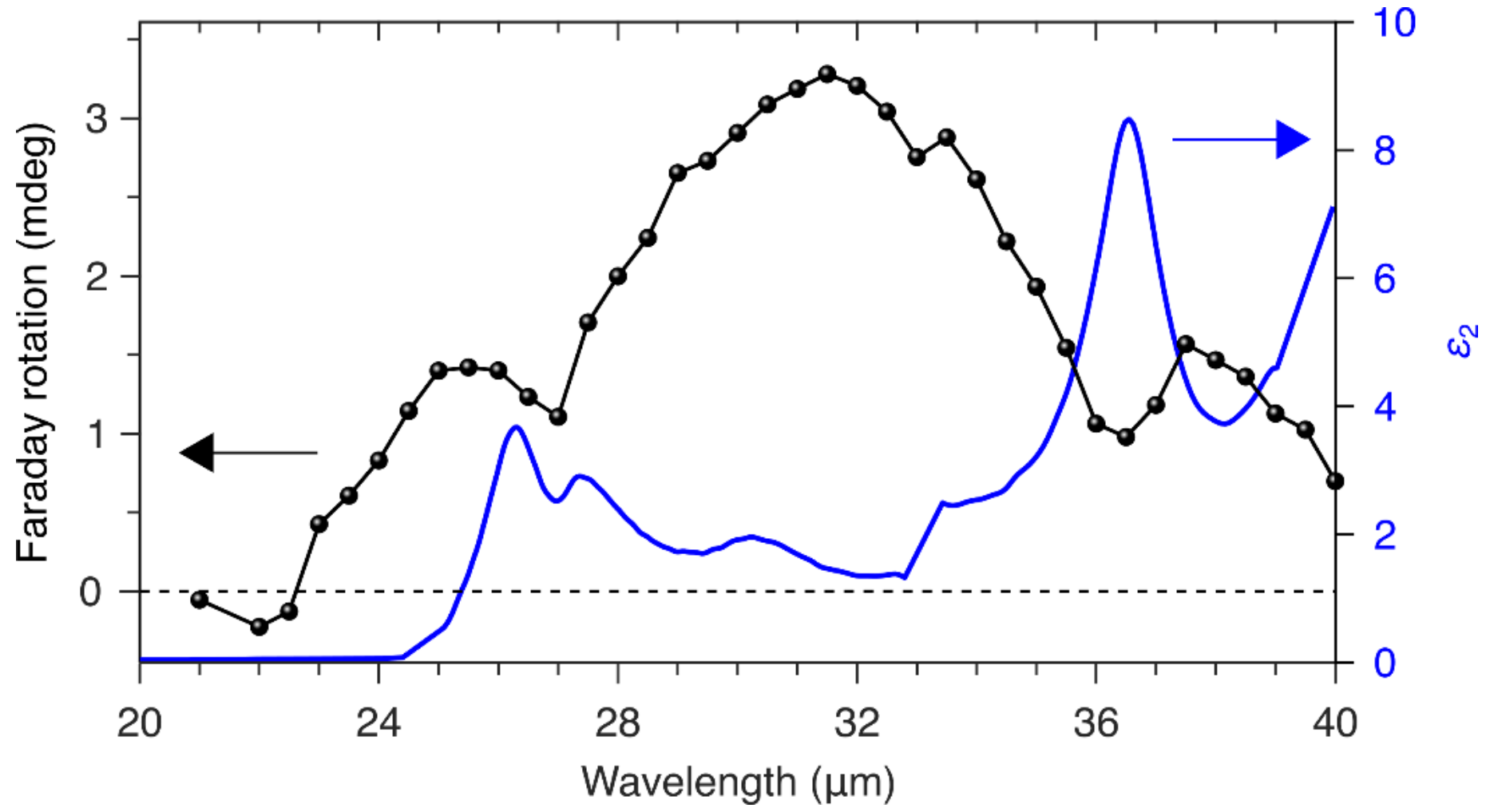

**Supplementary Fig. S8 | Spectral response near the higher-energy crystal-field excitation.** Expanded comparison of the normalized Faraday rotation spectrum (black symbols) and the optical absorption spectrum obtained from infrared ellipsometry (blue curve). The Faraday rotation amplitudes were extracted by fitting time-resolved pump–probe traces measured independently at each excitation wavelength. A broad enhancement of the Faraday rotation is observed in the vicinity of the higher-energy crystal-field absorption feature.

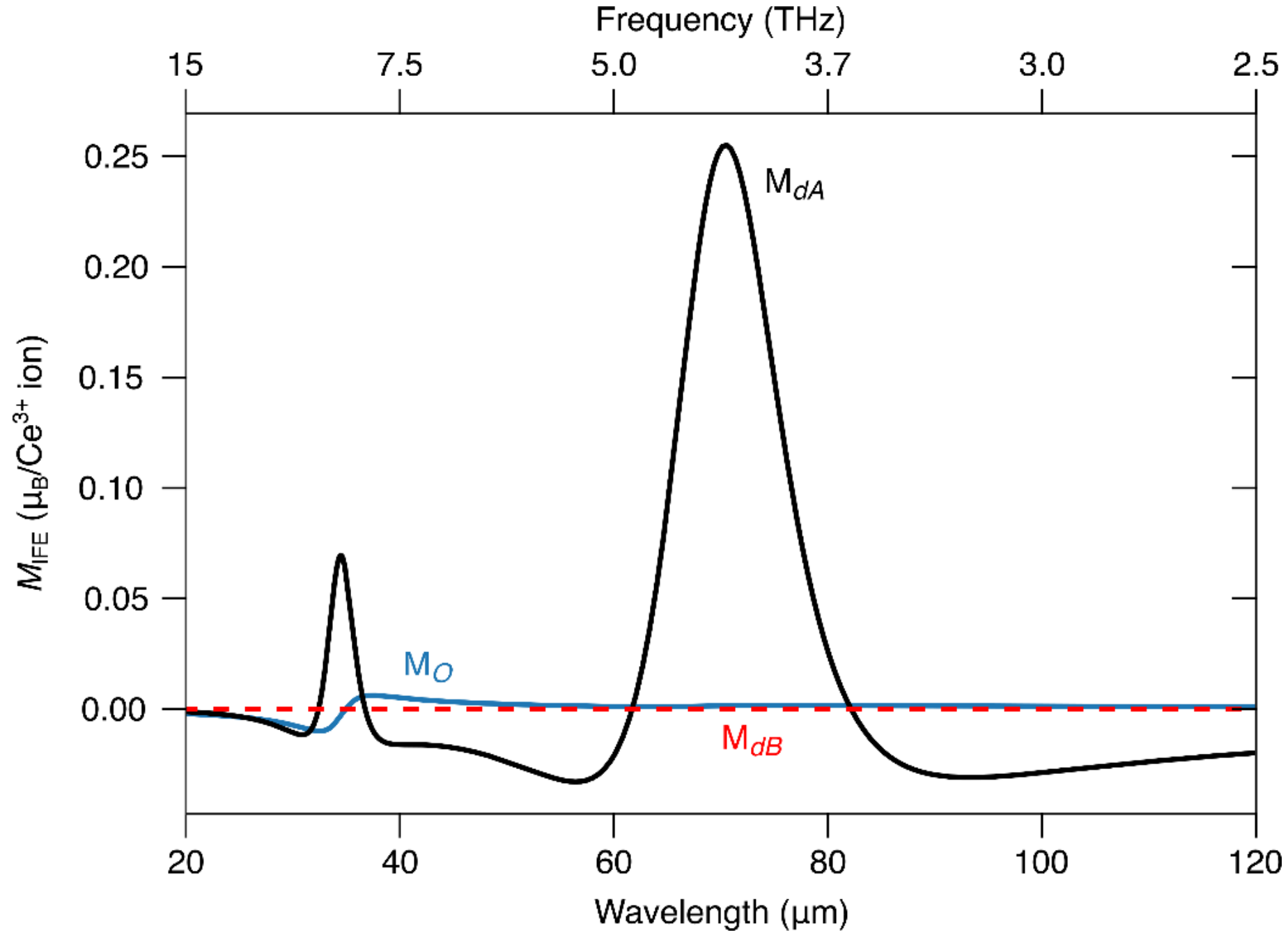

**Supplementary Fig. S9 | Decomposition of the crystal-field inverse Faraday effect.** Calculated spectral dependence of the THz-driven magnetization separated into the different contributions of the inverse Faraday effect stemming from $M_o$ (blue), $M_{dA}$ (black), and $M_{dB}$ (red).

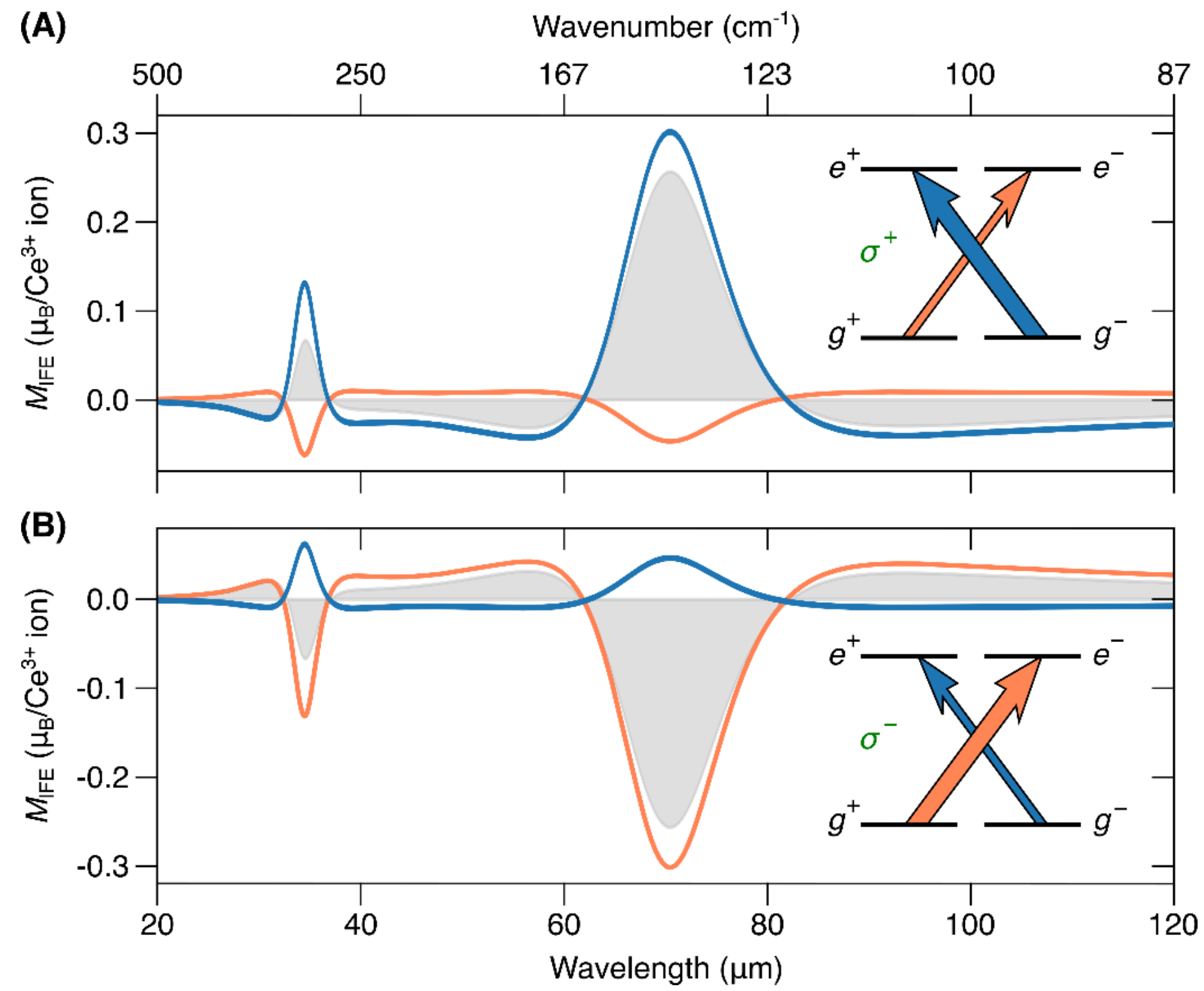

**Supplementary Fig. S10 | Microscopic origin of the sign reversal in the crystal-field inverse Faraday effect in $CeF_3$.** Calculated spectral dependence pf the THz-driven magnetization for **(A)** $\sigma^+$ and **(B)** $\sigma^-$ circularly polarized excitation. Blue and orange curves show the individual contributions from the helicity-allowed $g^- \rightarrow e^+$ and $g^+ \rightarrow e^-$ transitions, respectively, while the grey shaded area represents the total induced magnetization. Insets schematically illustrate the corresponding optical transitions, with the arrow thickness indicating the relative electric-dipole coupling strength.

**Supplementary Table 1 | Classification of optical phonon modes in $CeF_3$.** Best fit for the $E_u$ optical phonon modes in $CeF_3$, measured by infrared ellipsometry for in-plane polarization at 10 K. Their total number $N = 11$ is consistent with the group-theoretical vibrational analysis for the trigonal $P\bar{3}c1$ space group [S6]; $\omega_0$ [$\lambda_0$], $\Delta\varepsilon$ and $\gamma$ are their resonance frequency [central wavelength], contribution to the static permittivity and line width, respectively, according to the Lorentz oscillator model.

| $N$ | $\omega_0$ ($cm^{-1}$) [$\lambda_0$ (μm)] | $\Delta\varepsilon$ | $\gamma$ ($cm^{-1}$) |
|---|---|---|---|
| 1 | 99.0 [101.0] | 2.0 | 2.6 |
| 2 | 129.7 [77.1] | 0.2 | 2.7 |
| 3 | 164.3 [60.9] | 3.7 | 7.4 |
| 4 | 176.2 [56.8] | 0.2 | 7.8 |
| 5 | 197.8 [50.6] | 0.9 | 4.5 |
| 6 | 217.4 [46.0] | 0.8 | 11.9 |
| 7 | 245.4 [40.8] | 0.6 | 22.7 |
| 8 | 273.6 [36.6] | 0.3 | 10.9 |
| 9 | 330.5 [30.3] | 0.2 | 36.9 |
| 10 | 364.2 [27.5] | 0.1 | 15.1 |
| 11 | 380.8 [26.3] | 0.1 | 14.5 |

**Supplementary Table 2 | Crystal-field states of $CeF_3$.** The energies and wavefunctions of the crystal-field states of $CeF_3$, taken from Ref. [S8].

| Crystal-field state | Energy ($cm^{-1}$) | Wavefunction |
|---|---|---|
| $g^{\pm}$ | 0 | $0.7635\|\pm 5/2> + 0.6449\|\pm 1/2> - 0.0348\|\mp 3/2>$ |
| $e_1^{\pm}$ | 142 | $0.5348\|\pm 5/2> - 0.6615\|\pm 1/2> - 0.5259\|\mp 3/2>$ |
| $e_2^{\pm}$ | 290 | $0.3621\|\pm 5/2> - 0.3829\|\pm 1/2> - 0.8499\|\mp 3/2>$ |

## SUPPLEMENTARY REFERENCES